\documentclass[a4paper,12pt]{article}
\usepackage{amsfonts}
\usepackage{amsmath}
\usepackage{amstext}
\usepackage{amsthm}
\usepackage{xspace}
\usepackage[dvips]{graphicx}

\newcommand\ac{\`a\xspace}

\newtheorem{theorem} {Theorem}
\newtheorem{theorem*}{Theorem}
\newtheorem{prop*} {Proposition}
\newtheorem{lemma*}{Lemma}
\newtheorem{lemma}{Lemma}

\theoremstyle{definition}
\newtheorem{definition*}{Definition}
\newtheorem{cor*}{Corollary}
\newtheorem{rem*}{Remark}
\theoremstyle{remark}

\newtheorem{dim*}{\bf Dimostrazione }

\newtheorem{guess*}{\bf Osservazione}

\date{}
\input psfig.sty
\begin{document}
\title{Vortex dynamics in evolutive flows: a weakly chaotic
phenomenon} \author{Jacopo Bellazzini$^1$, Giulia Menconi$^2$,
Massimiliano Ignaccolo$^3$,\\
 Guido Buresti$^1$ and Paolo Grigolini $^{3,4,5}$} \maketitle
\begin{center}
\small $^{1}$ Dipartimento di Ingegneria Aerospaziale \\
\small Universit\ac di Pisa\\
  \small Via G. Caruso 56100 PISA -Italy\\ \small $^2$
Dipartimento di Matematica Applicata\\ \small Universit\ac di Pisa\\
\small Via Bonanno Pisano 25/b 56126 PISA - Italy\\ \small
$^{3}$ Center for Nonlinear Science\\ \small University of North
Texas\\ \small P.O. Box 311427, Denton, Texas 76203-1427\\ \small
$^{4}$ Dipartimento di Fisica\\ \small Universit\`{a} di Pisa and INFM,
\small Via Buonarroti 2, 56127 Pisa, Italy \\
$^{5}$Istituto di Biofisica del Consiglio Nazionale delle\\
Ricerche, Area della Ricerca di Pisa, Via Alfieri 1, San Cataldo,\\
56010,\\
Ghezzano-Pisa, Italy 

\end{center}
\begin{abstract}
We study the vortex dynamics in an evolutive flow. We carry out the
statistical analysis of the resulting time series by means of the
joint use of a compression and an entropy diffusion method. 
This approach to complexity makes it possible for us to establish that the time series emerging from the wavelet analysis of the vortex dynamics is a weakly chaotic process driven by a hidden  dynamic process. The complexity of the hidden driving process is shown to be located at the border between the stationary and non-stationary state. 
\end{abstract}
  \section {Introduction} In the last few years an increasing interest
has been devoted to the study of weak chaos, which has become for
many authors the paradigm itself of complexity. The characterization
of weak chaos, as it emerges in the context of deterministic chaos from
the work of Zaslavsky et al. \cite{zaslavsky}, implies the existence
of a phase space where regular regions coexist with random regions.
Some authors, see for instance Ref. \cite{wang}, propose a dynamic
condition where randomness and order are simultaneously present. The
latter is a property of systems, usually organized systems, where the
clock and coin flip paradigms apply at the same time
\cite{crutchfield}. Indeed, it is plausible that a condition of total
randomness is incompatible with the self-organization that is expected
to play a key role in such dynamical systems \cite{bak}.

In this sense, dynamic models of the same kind as the Manneville map
\cite{manneville} might be considered to be representative of the
complexity of living systems, as suggested, for instance, in the work
of Refs.\cite{Grigolini,giuliauno}.  The Manneville map has been
discussed as a model of intermittency years ago by Gaspard and Wang
\cite{GW} and more recently rigorous proofs and improvements of the
above results have been obtained by Bonanno \cite{claudio} and
Galatolo \cite{gal3}.

  We recall that the Manneville map $f(x)=x+x^z$ (mod. 1) is
characterized by a parameter $z>1$, which controls the frequency of
the random events. We can consider the waiting time between two
consecutive random events as time of sojourn in a condition of
order. The corresponding distribution of waiting times is an inverse
power law with index $\mu = z/(z-1)$. The condition $z = 2$ is the
border between the region where the waiting time distribution has a
finite mean ($z<2$) and the region where this mean waiting time is
infinite ($z> 2$). This latter behavior has been named {\it sporadic
dynamics}  \cite{GW}.

From a statistical point of view, the condition $z <2$ does
not mean ordinary statistical mechanics. In fact, the region $\frac 3
2\leq z<2$ is characterized by waiting time distribution with
divergent second moment and for this reason it was shown to be the
dynamic source of L\'{e}vy statistics \cite{anna,massi}. This form of
statistics led the authors of Ref.\cite{nature} to coin the term of
\emph{strange kinetics} to point out the anomalous nature of the
diffusion process emerging from this condition. The same authors
considered the strange kinetics as a synonym of weak chaos.  

Moreover, for a better understanding of the balance between order and
randomness in the Manneville dynamical systems, it is suitable to take
also the Kolmogorov-Sinai (KS) entropy \cite{ks} into account. In the
region $1<z<2$, the KS entropy is positive. The entropy vanishes when
the parameter $z$ enters the region $z\geq 2$, therefore generating a
non-stationary condition.  However, the characterization of weak chaos
by means of non-ordinary statistics is not sufficiently fine for our
purposes,
since it applies to both the case when the KS entropy is finite and to
the case when the KS entropy vanishes. Hereafter, we shall adopt the
following definition of weakly chaotic systems: systems whose KS entropy
is zero but which are not periodic. Therefore, following our
definition, the Manneville map is weakly chaotic only when the
parameter $z$ lies in the region $z\geq 2$.

Nevertheless, since our definition of weakly chaotic dynamical system
is focused on a vanishing KS entropy, we have to use an even  finer indicator
to classify the different types of weak dynamics. For instance, the
logistic map at the Feigenbaum point represents a weakly chaotic
scenario that arises from a period-doubling cascade. The authors of
\cite{mb,montangerofronzoni} showed that the logistic map at the chaos
threshold turns out to be mildly chaotic and is not an intermittent
form of weak chaos like the sporadic randomness defined by the
Manneville-like systems with $z>2$. In order to sketch the main
differences between the two kinds of dynamics, we can consider the quantity of
information which is necessary to describe $n$ steps of a typical
orbit. In the case of the intermittent Manneville-like systems with
$z>2$, the information grows as a power law $n^{\alpha}$ (where the
exponent $\alpha$ is $1/(z-1)$), while in the logistic map at the chaos
threshold it grows as the logarithm of $n$ (see Section 4 and
Refs. \cite{claudio,mb,licatone,gal3}).

The aim of this paper is to study a fluid dynamical process in the
outer shear layer of a coaxial jet flow: to be precise, the {\it
roll-up} and {\it pairing} of vortices. We reach the conclusion that
the observed dynamics is an example of weak chaos, driven by a hidden
intermittent process very well described by an idealized version of
the Manneville model.  In order to obtain our result we used three
distinct techniques. The first one is the wavelet-Hilbert analysis
\cite{invento}. The second one is the Diffusion Entropy (DE) method
\cite{deuno, giuliauno} and, finally, the third one is the Computable
Information Content (CIC) method of
Refs. \cite{ChaSolFra,licatone}. This is the second example of a joint
application of the DE and CIC methods.  The first is given by the work
of Ref. \cite{giuliauno}, where the reader can find a detailed
discussion about the benefit stemming from the joint use of these two
methods.

\begin{figure}[h]\begin{center}
\includegraphics[width=10cm]{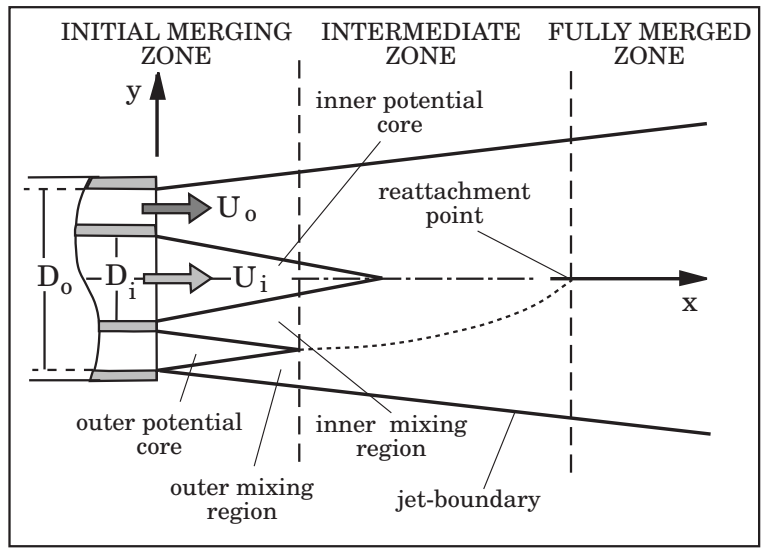}
\caption{\it illustration of the coaxial jet.}\label{getto}
\end{center}\end{figure}

As to the experimental apparatus used to acquire the data to study, the 
velocity signals are obtained in a coaxial jet flow (see
the scheme in Figure \ref{getto}), described in more detail by
Buresti et al. \cite{buresti}. The analyzed configuration has an
inner to outer jet diameter ratio $D_{o}/D_{i}=2.0$, $(D_{i}=50\ mm$),
a velocity ratio $U_{o}/U_{i}$=1.5 with $U_{o}=6\ m/s$ and wall
thickness $L=$5 $mm$. The axial and radial velocity components, $u$
and $v$, were measured by means of X hot-wires placed inside the inner
and the outer shear layers, at several axial positions downstream the
jet outlet. The measurements at each point were carried out three
times at a 5000 $Hz$ frequency for 20 seconds.

The outline of this paper is as follows. To make it as selfcontained 
as possible, in Sections 2, 3 and 4 we make a short review of the 
wavelet-Hilbert analysis, of the DE and of the CIC methods, 
respectively. In Section 5 we illustrate the results of these 
techniques applied to the vortex dynamics  under study in this 
paper. Finally, in Section 6 we discuss the results that were 
obtained.

\section{The Wavelet-Hilbert analysis}

The experimental velocity signals acquired in many flow fields are
cha\-ra\-cte\-rized by fluctuations induced by the passage or by the
oscillation of vortical structures. For instance, in the developing
shear layers of a coaxial jet the velocity signals are of increasing
complexity with increasing distance from the jet outlet, and multiple
dominating frequencies may be present, corresponding to fluctuations
with different degrees of modulation both in amplitude and in
frequency. These fluctuations are connected with the passage of
vortical structures produced by  the instability of the shear layers, 
and with their subsequent dynamics as well.  Therefore, signal analysis
procedures based only on conventional Fourier methods become largely
inappropriate and different techniques, providing the time variation
of the frequency and amplitude of the components present in the
signals, should be used. A classical demodulation technique, on which
the analysis procedures used in the present investigation are founded,
will now be briefly described. Any time-varying real signal, $x(t)$
may be represented in the form $$x(t)=A(t) \cos(\varphi (t))\ .$$ This
representation is not unique, but a so-called canonical pair may be
defined for the quantities $(A,\varphi)$ if we introduce the
associated complex analytic signal $Z_x(t)$ defined by $$Z_x(t)=A_x(t)
e^{i\varphi_x(t)}=x(t)+i x_H(t)\ ,$$ where $x_H(t)$ is the Hilbert
tranform of $x(t)$, given by $$x_H(t)=H[x(t)]=\frac 1 \pi
P\int_{-\infty}^{+\infty}\frac{x(\tau)}{t-\tau}\ d\tau\ .
$$ The canonical representation of $x(t)$ is then
$$x(t)=A_x(t)\cos[\varphi_x(t)], $$ hereby yielding the following
definition of instantaneous frequency of $x(t)$:
\begin{equation}\label{cinque} \nu_x(t)=\frac 1
{2\pi} \frac{d\varphi_x(t)}{dt}\ .
\end{equation} 
This definition, which is perfectly correct from a mathematical point
of view, acquires also a significant physical meaning when the analyzed
signal is asymptotic, i.e., a sinusoidal signal which is slowly
modulated in amplitude and frequency, $x(t)=A(t)\cos [\omega(t)\cdot
t+\varphi _0]$. In this case, provided the modulation frequencies
are sufficiently lower than the fundamental frequency of the signal,
the time variation of amplitude and frequency may be directly
recovered from the time variation of the modulus and of the phase
derivative of the associated analytic signal.

However, this procedure is applicable only if the signal is composed
of a single asymptotic component; conversely, if multiple components
are present, other techniques must be devised. One of these
\cite{Carmona} is based on the application of the continuous wavelet
transform and exploits the strict analogy existing between the
wavelet transform of a signal and the extraction of its associated
analytic signal when the used wavelet is a complex analytic function
(i.e. whose Fourier transform is zero for negative frequencies).
Indeed, the single modulated components present in the signal may be
extracted (provided they are asymptotic and sufficiently separated in
frequency) from the restrictions of the wavelet transform to the curves
where the wavelet modulus is maximum, which are called the {\it
ridges} of the transform.  More explicitly, if the signal is of the
form
$$x(t)=\Sigma_{n=1}^{N}x_n(t)=\Sigma_{n=1}^{N}A_n(t)\cos[\varphi_n(t)]=\Sigma_{n=1}^{N}A_n(t)\cos[\omega
_n(t)\cdot t+\varphi_{0n}] $$ and a Morlet wavelet
$\psi(t)=e^{i\omega_0 t}e^{-\frac{t^2}{2}}$ is used, then the
restrictions of the wavelet transform to the N ridge curves
$a=a_{rn}(t)$ (where the wavelet scale $a$ is proportional to the
inverse of the frequency) may be expressed as
$$W_{xn}(a_{rn}(t),t)=Corr(t)Z_{xn}+r(t), $$ where the term $Corr(t)$
is completely defined by the used wavelet and by the ridge values and
the residual $r(t)$ is negligible if the component is asymptotic.  It
is then possible to approximately obtain the analytic signals
associated to the single components and to estimate the relevant
frequency and amplitude modulation laws. However, some limitations
exist in the frequency modulation laws that may be detected using this
procedure \cite{Carmona}.  In particular, we note that the signals that do
not fit satisfactorily the asymptotic condition make the wavelet ridge
analysis much less accurate than the usual Hilbert
demodulation technique, see \cite{invento}. Consequently, a new
procedure was devised, which in a sense exploits the qualities of both
techniques and which will now be briefly described. The signal is
first transformed using a Morlet wavelet and the ridges are
approximately derived from the modulus maxima curves; in this
procedure, a high value of the central frequency of the wavelet
$\omega _0$ is used, in order to assure a high frequency resolution
and the reduction of the interference effects between adjacent
components. The wavelet maps are then filtered, neglecting the
coefficients outside a band around the dominating ridge, and an
inverse transform is applied.  The extracted signal may then be
subtracted from the original one and the procedure may be repeated
until all the detectable components are extracted. The Hilbert
transform technique is then applied to each component, in order to
obtain the required modulation laws.  In the latter step a further
filtering may be used, considering only the time intervals in which
the modulus of the associated analytic signal is higher than a given
threshold (in terms of a given percent of its mean value). This avoids
spurious large fluctuations of the instantaneous frequency and the
final statistical analysis may be restricted to those intervals of
time in which this frequency is physically meaningful, in the sense
that it may be confidently associated with fluctuations produced by a
fluid dynamical mechanism, like the passage or oscillation of
vorticity structures.  Obviously, in order to characterize their level
of chaoticity, the single components obtained through the wavelet
filtering procedure may also be analyzed with different techniques, as
for instance those described in the following sections.

\section{Diffusion Entropy Method}

This technique of analysis of time series was introduced originally 
in the article of Ref.\cite{deuno}, whose authors coined in fact the term 
Diffusion Entropy (DE). It became clear immediately 
afterwards \cite{giacomo}  that the DE method is an efficient method 
of establishing the scaling value if the scaling condition applies, 
regardless of the scaling origin, and even when the scaling 
condition, according to  the Generalized
Central Limit Theorem (GCLT) \cite{feller}, yields a divergent 
variance.  Given a discrete signal
$\big\{x_{i}\big\}_{i=1,N}$, we interpret it as a set of diffusion 
generating fluctuations. The collection of these fluctations yield a 
single diffusion trajectory. We can we create a set of many different
diffusion trajectories by means of moving windows of size $t$ with $1<t<N$.
We generate $N-t+1$ trajectories considering the sums:
\begin{equation}
y_{j}(t)=\sum_{i=j}^{i=j+t} x_{i}\ .
\end{equation}
Each diffusive trajectory can be thought of as the final position of a
walker which jumped for $t$ time steps. Let $p(y,t)$ be the
probability to be at position $y$ after $t$ time steps.
If the scaling condition applies to the asymptotic time limit, 
$p(y,t)$ is expected to fulfill  the following condition:
\begin{equation}
\label{scalingdefinition}
  p(y,t)=\frac{1}{t^{\delta}} F\left(\frac{y}{t^{\delta}}\right).
\end{equation}
Some of the processes of anomalous diffusion fit the prediction of 
the GCLT, thereby assigning to the function $F(y)$ the structure of a 
L\'{e}vy distribution with a diverging second moment. This is a 
problem for the techniques of analysis based on the observation of 
the second moment. It is not a problem for the DE, which evaluates the 
scaling parameter $\delta$ from  the Shannon entropy of the 
probability $p(y,t)$:
\begin{equation}
\label{shannonentropy}
S(t)=-\int_{-\infty}^{+\infty}p(y,t)\ln(p(y,t))dz.
\end{equation}
Indeed, it is straightforward to prove that when the scaling 
condition of Eq.(\ref{scalingdefinition}) holds, 
Eq.(\ref{shannonentropy}) yields $S(t)=A+\delta ln~t$, where $A$ is 
the Shannon entropy of the
generating function $F$. Thus, the scaling parameter $\delta$ is 
easily evaluated by plotting the Shannon entropy in a diagram with 
linear ordinate and a logarithmic abscissa. The Shannon entropy, as a 
function of time, becomes a straight line whose slope is the scaling 
parameter $\delta$.

It is important to point out that in the case where the time series
under study is a periodic process, the diffusion entropy,
starting from a small initial value, reaches a maximum, then it
regresses to the initial value, at a time corresponding to the period
of the process under study \cite{giuliauno}. Then, moving from this
value, the diffusion entropy increases again up to a maximum
which has the same value as the earlier maximum, and so on.  In the
case of the logistic map at the chaos threshold the diffusion entropy
exhibits a behavior very similar to that of a periodic process. The
maxima still lie on an horizontal line, but the regressions to the
minima are not complete as in the periodic case and occurr in a
disordered way \cite{giuliauno}.

 $\ $In the case of periodic processes the 
initial entropy increase is due to the fact that the moving window 
technique, adopted to create a large number of different diffusion 
trajectories, is equivalent to setting uncertainty on the initial 
condition. 

\section{Computable Information Content Method}
The second method of analysis used in this paper aims at establishing
a direct contact with Algorithmic Information Content (also known as
Kolmogorov complexity).
In this second method, the basic
notion is the notion of \textit{information}. Given a finite string
$s$ (namely a finite sequence of symbols taken in a given alphabet),
the intuitive meaning of quantity of information $I_{AIC}(s)$
contained in $s$ is the length of the shortest binary message from
which we can reconstruct $s$ (e.g. if we are working on a computer,
that binary message is a program $p$ that outputs $s$). This concept
is expressed by the notion of Algorithmic Information Content ($AIC$).
We limit ourselves to illustrating the intuitive definition of the
concept of $AIC$; for further details see \cite{cha} and
\cite{licatone} and related re\-fe\-ren\-ces.

By definition, the shortest program $p$ which outputs the string $s$ is a sort
of optimal encoding of $s$: the information that is necessary to
reconstruct the string is contained in the program. Unfortunately, this
coding procedure cannot be performed on a generic string by any
algorithm: the Algorithmic Information Content is a quantity which is not
computable by any algorithm (see Chaitin theorem in \cite{cha}).

Another measure of the information content of a finite string can also
be defined by a loss-less data compression algorithm $Z$ satisfying
some suitable optimality properties which we shall not specify here. 
Details are
discussed in \cite{licatone}. We can define the information content of
the string $s$ as the binary length of the compressed string $Z(s),$ namely,
\begin{equation}\label{definfoZ}
I_Z\left(  s\right)  =\left|  Z(s)\right|  .
\end{equation}

The advantage of using a compression algorithm lies in the fact that, this
way, the information content $I_{Z}\left(  s\right)$ turns out to be a
computable function. For this reason we shall call it Computable Information
Content $(CIC)$.

The notion of information is strongly related to chaos,
unpredictability and instability of the behavior of dynamical
systems. The KS entropy can be interpreted as the average measure of
information that is necessary to describe a step of the evolution of a
dynamical system.

We have seen that the information content of a string can be defined
either with probabilistic methods (following the Shannon theory) or
using the $AIC$ or the $CIC$. Similarly, also the KS entropy of a
dynamical system can be defined in different ways. The probabilistic
method is the usual one, the $AIC$ method has been introduced by
Brudno \cite{brud}; the $CIC$ method has been introduced in
\cite{galDCDS} and \cite{ChaSolFra}. So, in principle, it is possible
to define the entropy of a single trajectory of a dynamical
system. There are different ways to do this (see \cite{brud},
\cite{gal2}, \cite{GW}, \cite{claudio}, \cite{gal3}). In this paper,
we make use of a method which can be implemented in numerical
simulations. Now we shall describe it briefly.

Through the usual procedure of symbolic dynamics, given a discrete
infinite trajectory $\bar{x}=\big\{x_{i}=T^i(x_0)\big\}_{i\geq 0}$
drawn from the dynamical system $\left( X,\mu, T\right)$, we consider
a finite partition $\mathcal{P}=(R_{1},\dots,R_{l})$ of the dynamical
system. In a standard way, we associate a string
$\Phi_{\mathcal{P}}\left(\bar{x}\right) $ to the trajectory
$\bar{x}$. It is that $\Phi_{\mathcal{P}}\left(\bar{x}\right)
=(s_{0},s_{1},\dots,s_{k},\dots)$ if and only if
\[
\forall\ k\geq 0\ \ x_k\in R_{s_{_k}}\ , \quad\ \mbox{where }s_k\in\{1,\dots,l\} .
\]

We can define the {\it Information content} $I(\bar{x},\mathcal{P},n)$
of the trajectory $\bar{x}$ with respect to the partition $\mathcal{P}$
in the following way
$$I(\bar{x},\mathcal{P},n):=\tilde{I}(\Phi_{\mathcal{P}}\left(\bar{x}\right)
^{n})\ ,$$ where $ \Phi_{\mathcal{P}}\left(\bar{x}\right) ^{n}$ is the
string made of the first $n$ digits of the symbolic trajectory
$\Phi_{\mathcal{P}}\left(\bar{x}\right)$. The Information Content
$\tilde{I}$ can be measured either via $AIC$ or via $CIC$, so we have
just defined $I_{AIC}$ and $I_Z$ (respectively).

Let us assume that the compression algorithm $Z$ is {\it optimal} in
the sense of Ref. \cite{licatone}. We have the following results
(\cite{licatone}).

\begin{theorem}
  If $Z$ is an optimal coding, $(X,\mu ,T)$ is an ergodic dynamical
system and $\mathcal{P} $ is a measurable partition of $X$, then for $
\mu$-almost all trajectories $\bar{x}$ drawn from the dynamical
systems it holds:
\begin{equation*}
I_{Z}(\bar{x},\mathcal{P},n )=I_{AIC}(\tilde{x},\mathcal{P},n)=nh_{\mu
}(T,\mathcal{P} )+o(n)\ ,
\end{equation*}
where $h_{\mu }(T,\mathcal{P} )$ is the Kolmogorov entropy of $(X,\mu ,T)$ with
respect to the measurable partition $\mathcal{P} $.
\end{theorem}

\begin{theorem}
\label{cne} Given the dynamical system $(X,T,\mu)$, if the measure $\mu$
on $X$ is $T$-invariant, then, if $Z$ is an optimal compression
algorithm, for any measurable partition $\mathcal{P}$ it holds
\begin{equation*}
h_{\mu} (T,\mathcal{P})=\int_{X} \limsup\limits_{n\rightarrow 
+\infty}\frac{I_Z(\bar{x},\mathcal{P},n)} n \ d\mu =\int_{X}
\limsup_{n\rightarrow +\infty}\frac{I_{AIC}(\bar{x},\mathcal{P},n)} n \ d\mu\ .
\end{equation*}
\end{theorem}

Let us set $\limsup_{n\rightarrow +\infty}\frac{I(\bar{x},\mathcal{P},n)}
n$ be the complexity $K(\tilde{x},\mathcal{P})$ of the trajectory
$\tilde{x}$ with respect to the partition $\mathcal{P}$. Theorem \ref{cne} shows 
that if a system has an invariant measure, its
entropy with respect to a given partition can be found by averaging
the complexity of its orbits over the invariant measure. Then, the
entropy may be alternatively defined as the average orbit complexity.
However, if we fix a single point, its orbit complexity is not yet well
defined because it depends on the choice of a partition. We chose to
get rid of this dependence by considering only a particular class of
partitions and define the orbit complexity of a point as the supremum
of the orbit complexity over that class.

Let $\beta_{i}$ be a family of measurable partitions such that $
\mathrel{\mathop{\lim}\limits_{
i\rightarrow\infty}}\,diam(\beta_{i})=0$. If we consider the quantity
$\mathrel{\mathop{\lim\sup
}\limits_{i\rightarrow\infty}}K_{Z}(\bar{x},\beta_{i})$ we have that the
following lemma holds (for the proof, see \cite{licatone}).

\begin{lemma}
\label{1122} If $(X,\mu,T)$ is compact and ergodic, $Z$ is optimal, then for
$\mu$-almost all points $x_0\in X$, for the trajectory $\bar{x}$ of starting
point $x_0$ it holds $$ \mathrel{\mathop{\limsup
}\limits_{i\rightarrow\infty}}K_{Z}(\bar{x},\beta_{i})=
\mathrel{\mathop{\lim\sup }
\limits_{i\rightarrow\infty}}K_{AIC}(\bar{x},\beta_{i})=h_{\mu}(T)\ .$$
\end{lemma}

Therefore, this lemma, under suitable conditions on the dynamical
system, on the partitions and on the compression algorithm, permits to
define almost everywhere the {\it complexity of the orbit} $\bar{x}$
of the starting point $x_0$ as $\mathrel{\mathop{\lim\sup
}\limits_{i\rightarrow\infty}}K_{Z}(\bar{x},\beta_{i})=
\mathrel{\mathop{\lim\sup }
\limits_{i\rightarrow\infty}}K_{AIC}(\bar{x},\beta_{i})$. This
quantities asymptotically approach the KS entropy of the system.
Indeed, the asymptotic behavior of $I(x,\mathcal{P},n)$
gives an invariant of the dynamics which is finer than the KS entropy and is
particularly relevant when the KS entropy is null.

It is well known that the KS entropy is related to the
instability of the orbits. The exact relations between the KS entropy and the
instability of the system are given by the Pesin theorem. We shall recall
this theorem in the one-dimensional case. Suppose that the average rate of
separation of nearby starting orbits is exponential, namely,
\[
\Delta x(n)\simeq\Delta x(0)2^{\lambda n}\;\;\;\mbox{for }\Delta x(0)\ll 1,
\]
where $\Delta x(n)$ denotes the distance at time $n$ of two points
initially at distance $\Delta x(0)$. The number $\lambda$ is called
Lyapunov exponent; if $\lambda>0$ the system is unstable and $\lambda$
can be considered a measure of its instability (or initial data
sensitivity). The Pesin theorem implies that, under some regularity
assumptions, $\lambda$ equals the KS entropy.

In weakly chaotic systems, the amount of information necessary to
describe $n$ steps of a trajectory is less than linear in $n$ and the
rate of separation of nearby starting orbits is less than exponential,
then the KS entropy is not sensitive enough to distinguish the various
examples of weakly chaotic dynamics since in every case the KS entropy
is zero. Nevertheless, using the ideas we illustrated above, the
relation between initial data sensitivity and information content of
the orbits can be extended to these cases.

An example of such a generalization is given in Ref. \cite{gal3}. In
the following, for the sake of brevity we only briefly sketch the
ideas underlying the results of Ref. \cite{gal3}, which are very deep
and detailed. Let us consider a dynamical system $([0,1],T)$ where the
transition map $T$ satisfies some constructivity properties (a
constructive map is a map that can be defined using a finite amount of
information), and the function $I(\bar{x},\mathcal{P},n)$ is defined using
the $AIC$ in a slightly different way than before (use open coverings
instead of partitions). Provided the speed of separation of nearby
starting orbits goes like $\Delta x(n)\simeq\Delta x(0)f(x,n)$, it has
been proved in Ref. \cite{gal3} that under suitable assumptions on
$f$, we have for almost all the points $x\in\lbrack0,1]$ 
\begin{equation}
\label{chaos}
I(x,\mathcal{P},n)\sim\log(f(x,n)).
\end{equation}

In the weakly chaotic case, the speed separation of nearby starting orbits
is less than exponential.  In particular, if we have power law 
sensitivity $\Delta
x(n)\simeq\Delta x(0)n^{p}$, the information content of the trajectory is
\begin{equation}I(x,\mathcal{P},n)\sim
p\log(n)\ .\label{utile}\end{equation}

If we have a stretched exponential sensitivity $\Delta
x(n)\simeq\Delta x(0)2^{\lambda n^{p}}$, $p<1$ (e.g. in the
Manneville family with $z>2$) the information content of the orbits will
increase with the power law:
\begin{equation}
I(x,\mathcal{P},n)\sim n^{p} .
\label{utile2}
\end{equation}

Since we have shown that the analysis of $I(\bar{x},\mathcal{P},n)$
gives useful information on the underlying dynamics and since
$I(\bar{x},\mathcal{P},n)$ can be defined through the $CIC$
methodology, it turns out that it can be used to analyze experimental
data using a compression algorithm which is efficient enough and which
is fast enough to analyze long strings of data. For the numerical
experiments, we have used a compression algorithm called $CASToRe$,
illustrated in an earlier publication \cite{ChaSolFra}. The name of
the algorithm is an acronym meaning {\it Compression Algorithm,
Sensitive to regularity}; the heuristic motivations for this name are
explained in \cite{licatone}. Therefore, in the following the
algorithm $Z$ will be the algorithm $CASToRe$.

\section{Results}\label{resu}

As a significant earlier work on real fluid dynamical processes, we
have in mind the pioneering work of Solomon, Weeks and Swinney
\cite{swinney}, who studied chaotic transport in a laminar fluid flow
in a rotating annulus by tracking large numbers of tracer particles
for long times. The remarkable result of that paper is the detection
of L\'{e}vy statistics. This means that these authors discovered
strange kinetics. Our approach is slightly different. We do not adopt
the Lagrangian approach to chaotic transport of Ref. \cite{swinney},
but rather the Euler perspective: we measure the velocity fluctuations
in a suitable region of the flow.

In a jet flow, the shear layer between the so-called potential core
and the ambient fluid is subjected to an instability, which produces a
roll-up of the shear layer into vortical structures; moving
downstream, these structures grow and normally undergo a process of
pairing, through which two subsequent structures merge to form a
larger one. Depending on the geometrical and fluid dynamical
conditions, multiple pairings may occur before azimuthal instabilities
lead to a complete mixing of the flow at the end of the jet core. The
instability leading to roll-up is characterized by a dominating
frequency, while the pairing process leads to a doubling of the period
of passage of the structures, and then to the appearance of its
subharmonics. In the case of a coaxial jet (Figure \ref{getto}), the
situation is complicated by the presence of two cores and shear
layers, which give rise to different families of vortices whose degree
of interaction is a function of the ratios $\frac{D_o} {D_i}$ and
$\frac{U_o} {U_i}$. In the present case ($\frac{D_o}
{D_i}=2$,$\frac{U_o} {U_i}=1.5$), the stronger outer shear layer
dominates the initial development of the flow and its dynamics is not
far from that of a single jet  \cite{buresti}. In order to describe the
performance of the different signal processing techniques in providing
information on this dynamics, in the following we shall analyze the
axial velocity signal $u$ obtained from the measurements in the
position $\frac{x} {D_i}=1,\frac{y} {D_i}=0.96$, i.e. well inside the
outer shear layer.

\begin{figure}[h]
\begin{center}
\includegraphics[width=10cm]{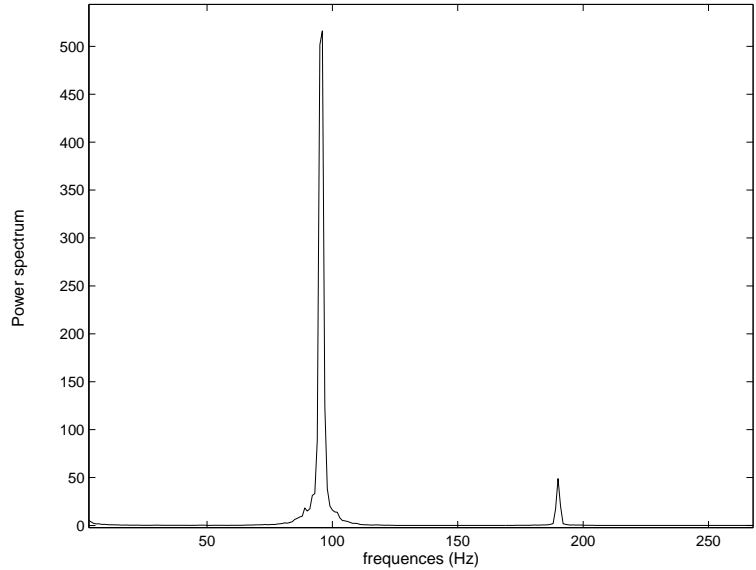}
\caption{\it Fourier power spectrum of the axial velocity $u$ at
position $x/D_i=1$ and $z/D_i=0.96$.}\label{spettro}
\end{center}\end{figure}

As shown in Figure \ref{spettro}, the Fourier spectrum of this signal
has two peaks, one at the fundamental frequency near $190$ Hz
(corresponding to a Strouhal number, based on $U_o$ and on the outer
shear layer momentum thickness at the jet outlet, $St=f\theta/U_o\sim
0.0125$), and a larger one at the subharmonic. This indicates that at
this position the pairing process is already in an advanced stage,
although not yet completed, as may be derived from the analysis of the
radial velocity spectrum (not shown here for the sake of brevity), in
which the peak at the fundamental frequency is still the dominating
one.

The two components of the $u$ signal were extracted using the wavelet
filtering procedure described above and we reconstructed the two
signals $u_a$ and $u_b$, associated to the harmonic ($u_a$, 190 $Hz$)
and to the subharmonic ($u_b$, 95 $Hz$) frequencies, respectively. The
two reconstructed signals $u_a$ and $u_b$ were then analyzed using the
different techniques described in the previous sections, starting from
the Hilbert demodulation procedure.

\begin{figure}[h]\begin{center}
\includegraphics[width=10cm]{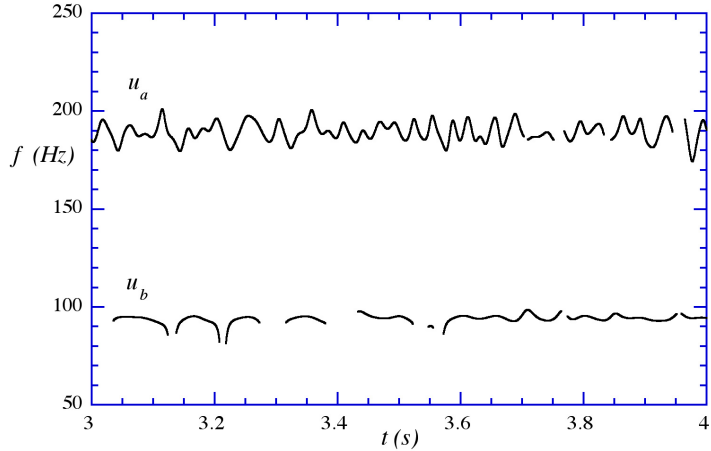}
\caption{\it Instantaneous frequencies of the components
$u_a$ and $u_b$ , as obtained using the wavelet-Hilbert procedure. }\label{tre}
\end{center}\end{figure}

Figure \ref{tre} shows a portion of the time variation of the
instantaneous frequencies of the two components derived from their
associated analytic signals. As already pointed out, to avoid
unphysical variations of this frequency due to the degeneracy of the
phase definition, the data are filtered considering only the time
intervals in which the modulus of the analytic signal is above a given
threshold ($50\%$ of the modulus mean value in Figure \ref{tre}) and
are then statistically analyzed. By this procedure, the mean values of
the two frequencies were found to be $188.80$ Hz and $94.47$ Hz, with
standard deviations of $4.8$ Hz and $0.93$ Hz, respectively. This
technique also allows us to obtain the time variation of the ratio
between the two instantaneous frequencies, whose mean value was found
to be $2.002$, with a standard deviation of $0.05$, thus confirming
that $u_b$ is the subharmonic of $u_a$. As can be seen from Figure
\ref{tre}, the instantaneous frequencies vary in time with an
irregular trend, and the same may be shown to be true for the modulus
of the analytic signal, which corresponds to the amplitude of the
original velocity component fluctuation. This irregular behavior is
most probably due to the fact that the shape, size, radial position
and translation velocity of the vortical structures are not exactly
constant in time. The nature of this irregularity will now be analyzed
through the DE and CIC methods.

\begin{figure}[h]
\begin{center}
\includegraphics[width=10cm,angle=270]{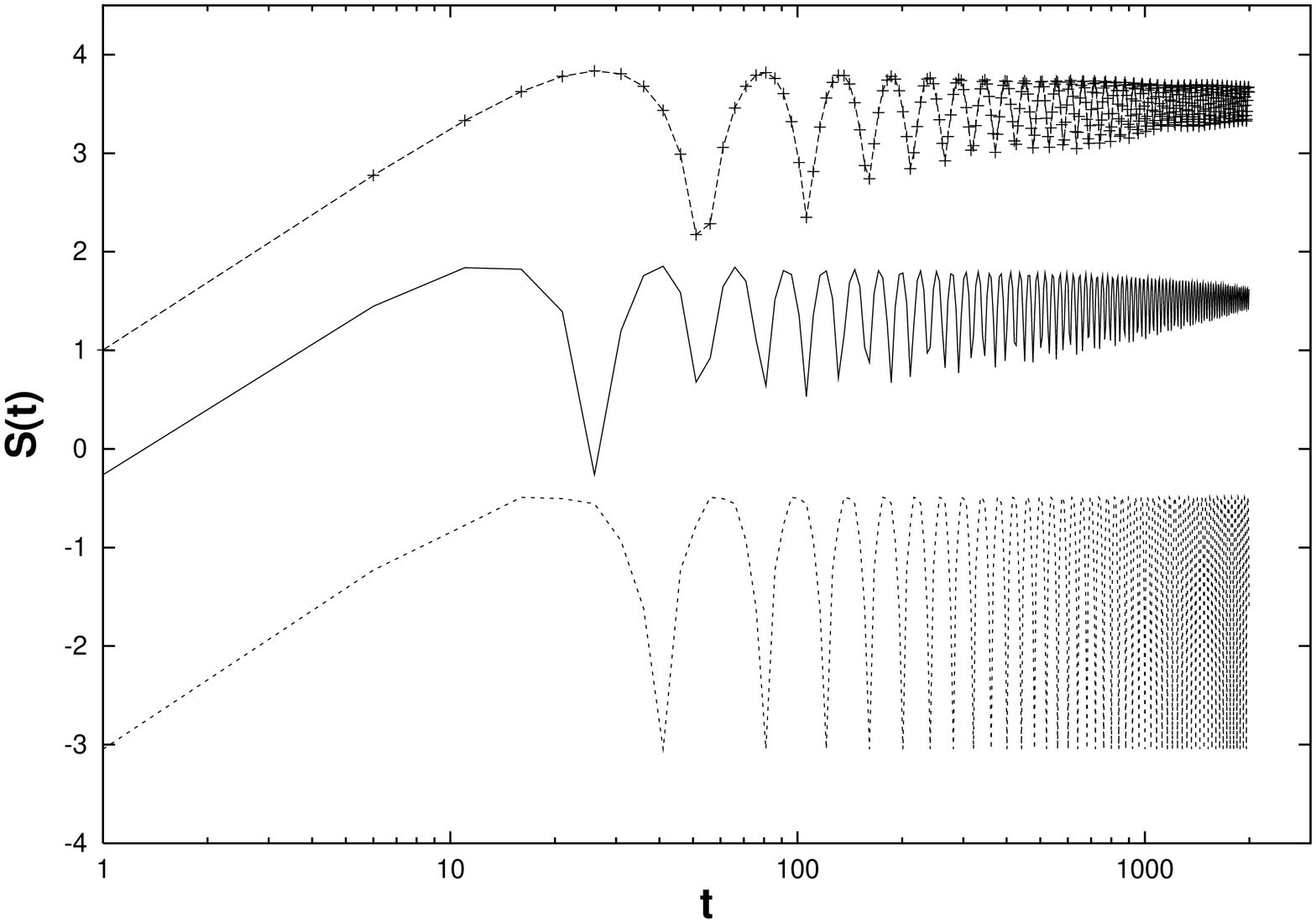}
\caption{\it results of the DE analysis on: $u_b$ (top graph, dashed
line), $u_a$ (mid graph, solid line) and a purely periodic signal
(bottom graph, dotted line). The scale is log-linear.}\label{cassandra}
\end{center}\end{figure}
First, we have studied the two reconstructed signals $u_{a}$ and
$u_{b}$ using the DE method.  The results are
illustrated in Figure \ref{cassandra} and, to make our discussion more
transparent, are compared to the result provided by the same method in the case
of a periodic signal.
The results of Fig. \ref{cassandra} deserve some detailed comments.  First of
all, we note that the two reconstructed signals yield a similar
behavior, with an initial transient followed by an oscillatory
behavior with the DE maxima lying on an horizontal line. This is
property shared by the periodic signal. It is e\-vi\-dent, however,
that the similarity with the periodic signal is not complete. Indeed,
in the case of the periodic signal the regressions of the DE to the
initial condition are complete, while in the case of the two
reconstructed signals they are not; in fact, the regression intensities tend
to decrease with time, suggesting that at infinite time the diffusion
entropy might become rigorously constant. This perfect localization
is similar to the slow occurrence of a collapse noticed year ago in
the case of a problem of quantum chaos \cite{bonci}. In that case the
dynamics under study was that of a $1/2$ dipole precessing about a
magnetic field while undergoing the perturbation stemming from the
harmonic motion of an oscillator. The entropy recursions to the
initial condition are complete if the Larmor frequency is very small
compared to oscillator frequency. In the case of re\-so\-nan\-ce, however,
chaos emerges and consequently incomplete regressions of decreasing
intensity were recorded.  All this suggests that the process under
study is neither periodic, nor quasi-periodic. In fact, in both cases
the entropy regressions to the initial condition would be complete
\cite{giuliauno,fronzonimontangero}.
 
In order to evaluate the degree of regularity of the two reconstructed
signals, we have estimated the dominant Lyapunov exponent of the two
time series \cite{rcdl}, which turned out to be zero for both
sequences $u_a$ and $u_b$. Therefore, thanks to the Pesin Theorem, we can
say that the vortex dynamics in both cases of roll-up and pairing has
null KS entropy. Together with the fact that the processes are not
ordered, we conclude that the vortex roll-up and pairing are weakly
chaotic phenomena.

\begin{figure}[h]\begin{center}
\includegraphics[width=8cm,angle=270]{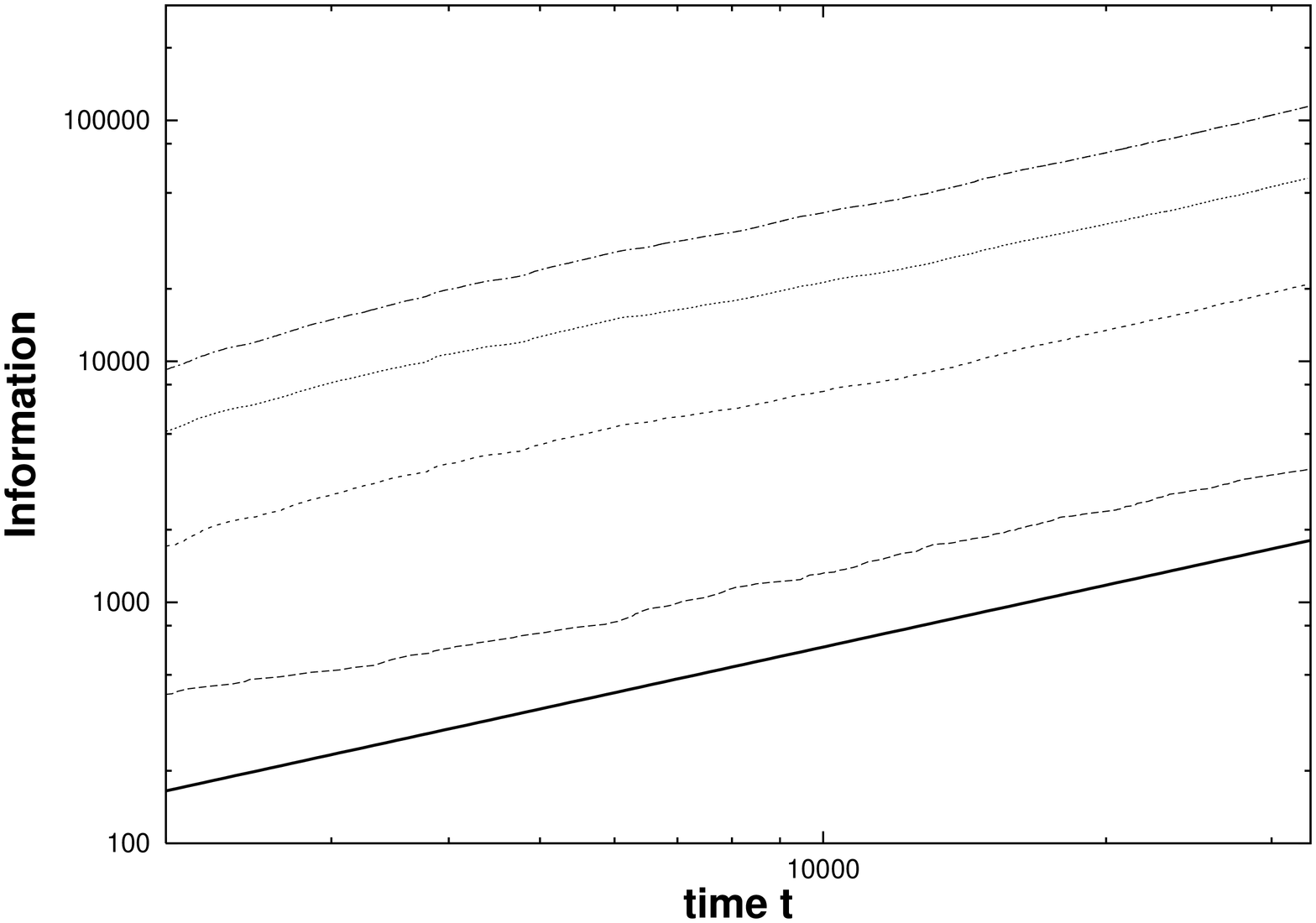}
\caption{\it The information content of the signal $u_b$ as a function
of the time length $t$, in a log-log representation. We show the
results corresponding to different partitions , al of them sharing the
same power law exponent $0.83$. From top to bottom: partition
$\mathcal{P}_{100}$, $\mathcal{P}_{32}$, $\mathcal{P}_8$,
$\mathcal{P}_2$.The bottom solid line is the analytical prescription
for the power-law exponent $0.83$.  }\label{cast}
\end{center}
\end{figure}

As a consequence, we classified the kind of complexity of the vortex
dynamics by investigating the behavior of the information content of
the reconstructed signals. We have translated the time series into
symbolic sequences by considering the interval $J_a=[\min\{u_a\},
\max\{u_a\}]$ (similarly, we defined $J_b$) and dividing it through
several uniform partitions $\mathcal{P}_k$ with $k$ subintervals where
$k\in\{2, 4, 8, 16, 32, 64, 100\}$.  As a result, the different
associated symbolic sequences gave the same power-law behavior for the
information content, which has been calculated by means of the
algorithm $CASToRe$:
\begin{equation}\label{legge}I_Z(n)\sim C n^{\alpha}\ .\end{equation}The exponent
$\alpha$ is $0.80$ for $u_a$ and $0.83$ for $u_b$; in both cases it
does not depend on the particular partition used. The constant $C$
depends on the partition $\mathcal{P}_k$ and increases with the
parameter $k$ (see figure \ref{cast}).  As illustrated in section 4,
the fact that the amount of information to describe $n$ steps of a
trajectory is less than linear in $n$ is in complete agreement with
the null KS entropy.

\begin{figure}[h]\begin{center}
\includegraphics[width=8cm,angle=270]{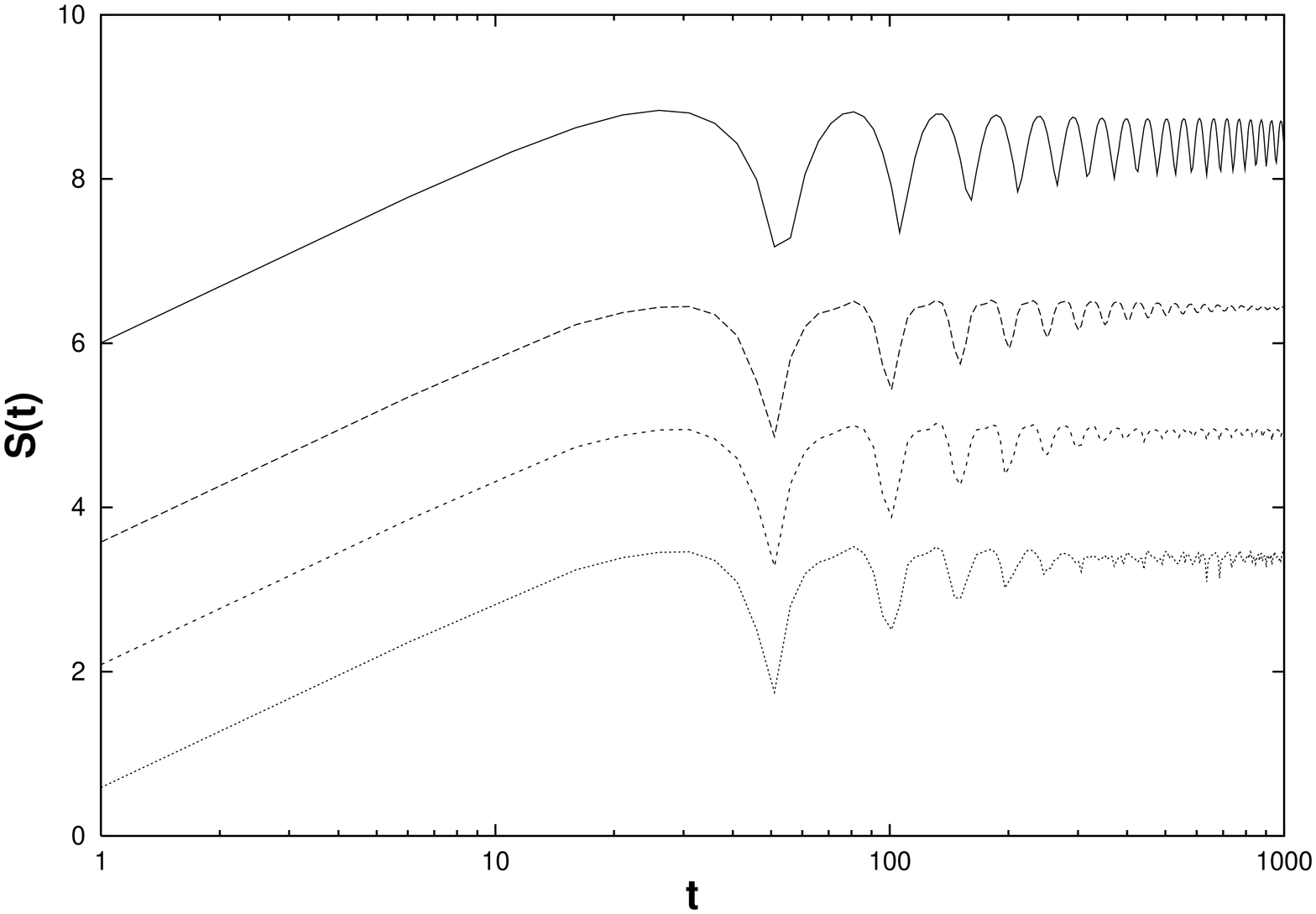}
\caption{\it result of the DE analysis on: $u_b$ (top graph, solid
line) and the simulated signals $\sigma_{\mu}(t)$ generated with different 
values
of the control parameter. From top to bottom: $\mu=3.5$, $\mu=2.45$, 
$\mu=1.90$.
}\label{estetica}
\end{center}
\end{figure}

At this stage, we have to address the challenging task of building up
a model reproducing the main properties of vortex dynamics found so
far. This model must explain the degree of complexity as it is indicated by the
parameter $\alpha$ ranging from $0.80$ to $0.83$, and at the same time
it has to be compatible with another crucial property such as the
two-peaks Fourier spectrum of Fig. 2. We propose a theoretical model,
resting on the basic idea that the experimental signals are driven by
a hidden intermittent process inspired to the Manneville map. We shall
use this model to reproduce both the signal $u_{a}$ and the signal
$u_{b}$. However, we shall give details only on the choices made to
reconstruct $u_{b}$. The prescriptions adopted to reconstruct the
signal $u_{a}$ are similar, and, for the sake of simplicity, will not be
illustrated.  First of all, we define the following waiting time
distribution:
\begin{equation}
\label{power}
\Psi(\tau) = \frac{(\mu-1)T^{\mu-1}}{(T+\tau)^{\mu}}\ .
\end{equation}
Note that this form is chosen as the simplest way to generate an
inverse power law at long times, without conflicting with the
normalization condition. In the case $\mu> 2$, the first moment of
this distribution is $T/(\mu -2)$, thereby showing that the parameter
$T$ keeps under control the mean value of this time distribution.  We
also define the set of frequency values $A_{\ell}=\{
\omega_1,....,\omega_{\ell} \}$ where the values $\omega_j$'s are
determined on the basis of the Hilbert transform approach used to
derive the results of Fig. \ref{cassandra}. More precisely, these values can be
considered to be a coarse-graining representation of the condition
illustrated in Fig. \ref{tre}. In the specific case of the signal $u_b$, we
have fixed $\ell=10$ and the frequencies $\omega_j$'s have been
determined by a uniform partition of the frequency centered at
($94.47$ Hz) with a width that is twice the standard deviation ($0.93$
Hz).

We now define the artificial signal as follows. We select a time $\tau
_1$ of the distribution of Eq.(\ref{power}), we call laminar phase
this time interval and we randomly select one of the frequencies of
the set $A_{\ell}$, say $\omega(\tau _1)$. We assume that the signal,
a sinuisoidal function of time, keeps this frequency throughout the
whole laminar region. At the end of this laminar region, we select
another number $\tau _2$ from the distribution $\psi(\tau)$ and another
frequency $\omega(\tau _2)$ from the set $A_{\ell}$, and so on. At any random
drawing of the pair $\{\tau_j, \omega(\tau _j)\}$, we must also select a
phase $\phi_j$. This latter choice is not arbitrary, but it is done in
such a way as to ensure that the resulting signal is continuous. In
conclusion, we get the following signal
\begin{equation}
\label{dynamicmodel}
\sigma_{\mu}(t) =  \sin(\omega_{\mu}(t)t + \phi_{\mu}(t)),
\end{equation}
where the subscript $\mu$ indicates the dependence of this artificial
signal on the random distribution of Eq. (\ref{power}), with a fixed value of
$\mu$. This special way of sewing a laminar region to the next has been
dictated by the need of reproducing the experimental spectrum of
Fig. 2. Our numerical calculations allowed us to assess that a
different choice, based on the abrupt frequency jump , with a fixed
phase, and consequently with an abrupt jump of the signal
$\sigma_{\mu}(t)$, yields a continuous Fourier spectrum, thereby
dramatically departing from the behavior illustrated by Fig. 2.

 The inverse power law distribution given by (\ref{power}) is an
idealization of the Manneville map mentioned in the
introduction. Nevertheless, the information content growth of the
instantaneous frequency $\omega_{\mu}(t)$ is the same as that of a
trajectory drawn from a Manneville map with driving parameter
$z=\frac{\mu}{\mu-1}$. Therefore, the three types of statistics
illustrated in the Introduction become: $\mu>3$, ordinary statistics;
$3>\mu>2$ strange kinetics in the sense of Zaslavsky \cite{zaslavsky}
and $\mu<2$ sporadic dynamics in the sense of Gaspard and Wang
\cite{GW}. The sewing process adopted to reproduce the Fourier
spectrum of the real signal has the effect of hidding this complexity
so as to make the DE analysis insensitive to the complexity of this
hideen process. This is clearly illustrated in Fig. 6, where we see
that the artificial sequences with three different degrees of hidden
complexity, ranging from weak chaos ($\mu = 1.90$) to ordinary
statistical mechanics ($\mu = 3.5$) through a condition of strange
kinetics ($\mu = 2.45$), result in the same time behavior of $S(t)$.
We note that, in spite of a lack of sensitivity to the complexity of
the hidden driving process, Fig. \ref{estetica} proves that there is a
good agreement between the DE of the reconstructed signal $u_b$ and
the DE of three simulated sequences generated with different driving
parameter $\mu$.  Therefore, Eq.(\ref{dynamicmodel}), with the random
choice of waiting times prescribed by Eq.(\ref{power}), is a plausible
model of the turbulent process under study.

\begin{figure}[h]\begin{center}
\includegraphics[width=8cm]{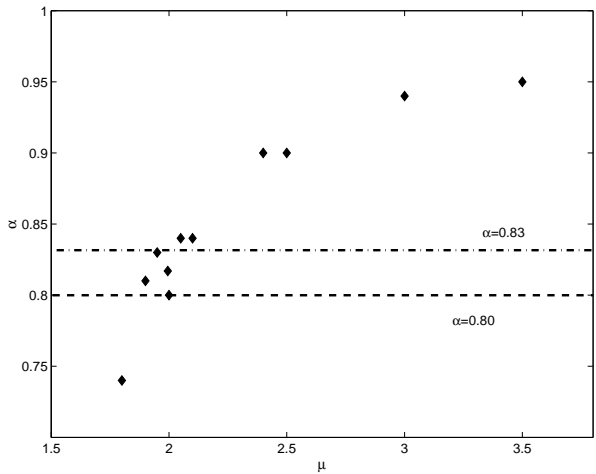}
\caption{The index $\alpha$ denoting the complexity of the signal
$\sigma_{\mu}(t)$ of Eq. (10) as a function of the parameter $\mu$,
denoting the complexity of the hidden driving process.}\label{alpha}
\end{center}
\end{figure}

In conclusion, while the complexity of the reconstructed signals is
measured with the compression algorithm and turns out to be given by
Eq.(\ref{legge}) with $\alpha=0.80$ in the case of $u_a$ and
$\alpha=0.83$ in the case of $u_b$ (see Fig. \ref{cast}), the
complexity of the driving process $\omega(t)$ remains hidden. At the
moment we do not have any analytical expression to relate the
information content index $\alpha$ to the driving parameter $\mu$ of
Eq.(\ref{dynamicmodel}). Thus, it is impossible to determine
theoretically the degree of complexity of this driving process. We
must rest on a numerical treatment, based on applying the compression
algorithm to artificial sequences with different values of $\mu$.  For
this purpose, we have turned the artificial time series generated by
Eq.(\ref{dynamicmodel}) into several symbolic sequences by means of a
uniform partition $\mathcal{P}_2$. For all the values of $\mu$ adopted
in the numerical simulations, the resulting information content of
associated symbolic series turned out to be Eq.(\ref{legge}), with
$\alpha<1$, which is equivalent to the sporadic randomness of Gaspard
and Wang \cite{GW}. The dependence of $\alpha$ on $\mu$ is pictured in
Figure \ref{alpha} and turned out to be essentially monotonic. The
numerical function $\alpha(\mu)$ crosses the horizontal strip
$0.80\leq\alpha\leq 0.83$ at $\mu=2\pm 0.1$. Thus the complexity of
the driving stochastic process is measured, with a good approximation,
by equation (\ref{dynamicmodel}) with $\mu = 2$.

\section{Concluding remarks}
The interest of this paper rests on the discovery of the hidden
complexity of the signals $u_{a}$ and  $u_{b}$, extracted from the
vortex dynamics by means of the wavelet analysis. The main properties
of these signals are satisfactorily reproduced by the artificial
signal of Eq.  (\ref{dynamicmodel}), obtained according to the
procedure described in Section \ref{resu}. This important aspect is illustrated
by Fig. \ref{alpha}. For this reason we devote these concluding remarks to 
further comments on this figure and we stress that from it two main
results emerge.

The first result is that the sewing process adopted to generate the
artificial sequence yields a sort of time dilution of randomness. The
driving signal for $\mu> 2$ is characterized a non-vanishing Lyapunov
coefficient and consequently by an Information Content growing
linearly in time. The time increase of the Information Content becomes
sub-linear only in the region $\mu < 2$. We see from Fig. \ref{alpha}
that the Information Content of the signal $\sigma_{\mu}(t)$ increases
sub-linearly in time throughout the whole range of $\mu$-values
considered, and remains probably sub-linear well beyond $\mu=3$, which
is the border between L\'{e}vy s and ordinary Gaussian
diffusion. According to the definition of weak chaos adopted in this
paper, we conclude that the signal $\sigma_{\mu}$ is an expression of
weak chaos regardless of the degree of hidden complexity. We do not
have available at the moment a theory expressing $\alpha$ as a
function of $\mu$. However, the numerical results of Fig. \ref{alpha}
indicate that $\alpha(\mu)$ is a monotonic function of $\mu$ with
$\alpha < 1$. The second result has to do with the detection of the
degree of hidden complexity, namely, the complexity of the driving
signal $\omega_{\mu}(t)$. The uncertainty on the complexity of the
signal $u$ is measured by the width of the strip between the straight
line $\alpha = 0.83$ and the straight line $\alpha = 0.80$. Thus, the
uncertainty on the complexity of the hidden signal can be determined
by observing the crossing between the experimental curve $\alpha(\mu)$
and the strip. We see from Fig. \ref{spettro} that the crossing occurs in a
sharp interval of $\mu$ values around $\mu = 2$. More precisely, the
complexity of the driving signal is measured by $1.9 < \mu <
2.1$. This makes us conclude that the complexity of the driving signal
lies at the border between weak chaos ($\mu < 2$) and strange kinetics
($2 < \mu < 3$. There are other complex processes lying very close to
this border (see, for instance, the solar flares of Ref. \cite{debbie} and the
earthquakes in California \cite{terremoti}). It would be interesting to assess
why the border between the stationary and the non-stationary states is
the basin of attraction of some complex systems.

\emph{Acknowledgment} P.G. gratefully acknowledges financial support
from ARO, through Grant DAAD19-02-0037.

\end{document}